\RequirePackage{fix-cm}
\documentclass[smallextended]{svjour3}
\smartqed
\usepackage{tikz}
\usepackage[labelsep=none]{caption}

\usepackage{amsfonts}
\usepackage{clrscode}
\usepackage{grffile}
\usepackage{graphicx}
\usepackage[noend]{algorithm2e}
\newtheorem{defn}[theorem]{Definition}
\newtheorem{lemmx}[theorem]{Lemma}
\newtheorem{prop}[theorem]{Proposition}

\newcommand{\prp}[1]{Proposition~\ref{p.#1}}
\newcommand{\secn}[1]{Section~\ref{s.#1}}

\newcommand{\fig}[1]{Figure~\ref{f.#1}}

\newcommand{\eq}[1]{(\ref{e.#1})}
\newcommand{\overx}{{\overline x}}
\newcommand{\overp}{{\overline p}}

\def\proofend{\hfill\mbox{${\vrule height6pt width6pt depth0pt}$\medskip}}

\def \proof{\noindent{\bf Proof.}\enspace}

\def\set#1{\{#1\}}

\def\R{\rlap{\rm I}\,{\rm R}}

\begin{document}
\bibliographystyle{spmpsci}

\title{Unbalancing Binary Trees}
\author{Matthew L. Ginsberg}
\institute{M.L.~Ginsberg \at
  Eugene, OR~~97405 \\
  {\sf https://orcid.org/0000-0001-9343-354X} \\
  \email{mlginsberg@gmail.com} }
\maketitle
\begin{abstract}
\begin{quote}
  Assuming Zipf's Law to be accurate, we show that existing techniques
  for partially optimizing binary trees produce results that are
  approximately 10\% worse than true optimal.  We present a new
  approximate optimization technique that runs in $O(n \log n)$ time
  and produces trees approximately 1\% worse than optimal.  The
  running time is comparable to that of the Garsia-Wachs algorithm but
  the technique can be applied to the more useful case where the node
  being searched for is expected to be contained in the tree as
  opposed to outside of it.

\end{quote}
\end{abstract}

\section{Introduction}
\label{s.intro}

Binary trees are often used to store and index large amounts of data.
Many such trees are balanced in the sense that the left and right
children of any particular parent are similar in size.  A consequence
of this is that every node can be reached in at most $\log_2(n)$ steps
from the root of the tree, where $n$ is the tree's size.

While a bound on worst case performance is obviously desirable, a
bound on {\em average-}case performance may be more important in an
environment where large trees are searched often.  Note that in cases
where searches are frequent indeed, statistical information gathered
from past searches can be used to determine the probability $p(x)$
that a particular node $x$ in the tree is the target of a randomly
selected query.  If we denote the depth of a node $x$ by $d(x)$, the
problem of optimizing for average-case performance becomes that of
finding a tree structure $T$ for which the cost
\[c(T) = \sum_{x\in T} p(x)d(x)\]
is minimized, where $T$ is the set of all nodes in some particular
tree and $c(T)$ is the expected cost of a single lookup in the tree.

A variety of authors have considered this problem, and two approaches
have historically been taken.  Some researchers have focused on
finding a globally optimal tree, while others have focused on methods
that construct approximately optimal trees.  As one might expect, the
complexity of the approximate methods is lower than their exact
counterparts.  Knuth \cite{Knuth:binary} provides a reasonable summary
with focus on the construction of globally optimal trees, while
Nagaraj \cite{Nagaraj:binary} provides a somewhat broader description.

Finding a globally optimal tree in general takes time $O(n^2)$,
prohibitive given the size of modern search trees.  There are two
possible improvements:

\begin{enumerate}
\item Karpinski et.al.~\cite{Karpinski:optimal} have developed an
  $O(n^{1.6})$ algorithm if there is a fixed lower bound on the
  relative probability of any particular node in the sense that there
  is some fixed $\delta$ such that if $\set{v_1,\dots,v_n}$ is an
  ordering of the values in the tree, then the probability that a
  query is between $v_{i-1}$ and $v_{i+1}$ is always at least
  $\delta/n$.  This assumption of relative uniformity among the
  weights of the nodes is unlikely to be valid in practice because of
  Zipf's law \cite[and \secn{base}]{Zipf} and the fact that the
  harmonic sequence $\sum_{i=1}^n \frac 1i$ diverges.  In any event,
  $n^{1.6}$ is itself unlikely to be viable for large trees.
\item Garsia and Wachs \cite{Garsia-Wachs} and many successors
  \cite[and others]{Hu:binary,Karpinski:binary,Mehlhorn:optimal} have
  considered $O(n \log n)$ algorithms in the special case where it is
  known that the element $v$ being searched for is not in fact in the
  tree in question, and the goal is therefore to find two consecutive
  existing values $v_i,v_{i+1}$ with $v_i < v < v_{i+1}$.  It is
  perhaps not surprising that the complexity is lower in this case,
  since the placement of the internal nodes is less important if the
  search must proceed to the fringe of the tree in any case.  Once
  again, however, this work is of limited practical use in a setting
  where (for example) one is trying to make Internet search more
  efficient -- after all, when is the last time you did a Google
  search and got no results?
\end{enumerate}

A wide variety of authors have also considered fast (generally $O(n
\log n)$) methods for constructing nearly optimal binary trees.  The
three most popular ideas appear to be the following:

\begin{enumerate}
  \item {\bf Splay trees} were introduced by Sleater and Tarjan
    \cite{Tarjan:splay}.  The tree is initially constructed randomly,
    but every query moves its target to the top of the tree, hopefully
    causing frequent queries to have small depths.  Splay trees are
    intended to be adjusted online as queries are received, as opposed
    to depending on a knowledge of relative probabilities in advance.
  \item {\bf Treaps} were developed by Aragon and Seidel
    \cite{Aragon:treap,Seidel:treap} although the method is hinted at
    earlier by Mehlhorn \cite{Mehlhorn:nearly}.  In this approach, the
    binary tree is constructed by inserting points in probability
    order.  Although the average case complexity is within a constant
    factor of optimal, Mehlhorn points out that in the worst case, the
    tree produced may be quite far from optimal.  Mehlhorn argues that
    the method should be discarded for that reason.
  \item {\bf Weight-balanced binary trees} also appear in Mehlhorn's
    work \cite{Mehlhorn:nearly}.  In the construction of the tree or
    any subtree, the node at the root is the one that most evenly
    divides the aggregated probabilities of the residual nodes between
    the left and right subtrees.

    If $H$ is the entropy of the frequency distribution of the nodes
    in the tree, Mehlhorn shows that the cost $C$ of the tree
    constructed in this fashion is bounded below by $H/{\log 3}
    \approx 0.63 H$ and above by
    \[ 2+H / [1-\log(\sqrt 5 - 1)] \approx 1.44 H,\]
    and that these bounds apply to the optimal tree as well.  For
    $H\geq 14.5$, the lower bound was subsequently improved by De
    Prisco and De Santis \cite{DePS:lower} to
    \[H + H\log H - (H+1)\log(H+1) \geq H - \log H - 1 - \frac 1H \approx H-1.\]
\end{enumerate}

For treaps and weight-balanced trees, it can be shown that the average
case performance of the data structure in question is within some
constant factor of optimal.  (Indeed, for weight-balanced trees, the
cost is within a constant factor of optimal in every case.)  This is
suspected to be the case for splay trees as well, but this ``dynamic
optimality conjecture'' has not been proven.  The best result of this
general type appears to be due to Lecomte and
Weinstein~\cite{Lecomte:tango}, who show that Tango Trees
\cite{Demaine:tango} are within a factor of $O(\log(\log(n)))$ of
optimal.

In today's world, however, where significant computing resources are
devoted to finding objects on the Internet, even a pure constant
factor is arguably not good enough.  All of the above methods (except
Tango Trees) were developed in an environment where it was assumed
that tree searches would be approximately as common as
insertions.\footnote{Donald Knuth, personal communication.}  Such an
assumption is obviously no longer valid today; I (thankfully!)  search
the Internet far more frequently than I modify it in a publicly
accessible way.

Our goal in this paper is to describe a technique that unbalances
binary trees in a way that improves their average-case performance.
The technique that we will present can run in time $O(n)$, although
performance will be better as run time is permitted to increase, and
is guaranteed to improve the expected performance of the tree being
optimized.

We discuss our basic experimental framework and the baseline
corresponding to existing techniques in the next section, and our new
ideas are introduced in \secn{technique}.  Experimental results are
presented in \secn{exp}, and suggestions for future work appear in
\secn{next}.

\section{Baseline}
\label{s.base}

We will work with randomly generated binary trees of 100 to $10^7$
nodes.  In each case, we follow Knuth \cite{Knuth:binary} and assign
weights to the nodes using Zipf's Law~\cite{Zipf}, so that the
$n^{th}$ most popular node has relative probability $1/n$.  A variety
of other authors have concluded both through observation \cite[and
  others]{Breslau:zipf,Cunha:zipf} and theory \cite[and
  others]{Huberman:zipf} that Zipf's Law reasonably approximates the
relative number of accesses to any particular web site, and therefore
presumably the number of searches for a particular word or phrase on
the Internet as well.\footnote{Some authors suggest that the
  popularity of node $n$ should be proportional to $n^{-\alpha}$ for
  some $\alpha$ but it appears both that the results we present are
  not terribly dependent on the exact nature of the distribution in
  question and that $\alpha$ is close to one in any event.}

For each size of tree investigated, we consider five basic types of
trees: simple binary trees, treaps, weight-balanced trees, splay
trees, and trees that are provably optimal given the underlying weight
distribution.  For splay trees, we generate random but appropriately
distributed queries and repeatedly move the node being searched for to
the root of the tree.  The number of queries generated is $3n$, where
$n$ is the size of the tree.

Optimal orderings were only computed for trees of 50,000 nodes or
fewer.  Both the time and (more importantly) the memory needed by the
best known algorithm here \cite{Knuth:binary} are $O(n^2)$ where $n$
is the number of nodes in the tree, and $n>50000$ was simply too large
to be practical on the machine being used (a 32-core AMD Ryzen with
128GB of memory).

\begin{figure}
\begin{center}
  \includegraphics[width=\textwidth]{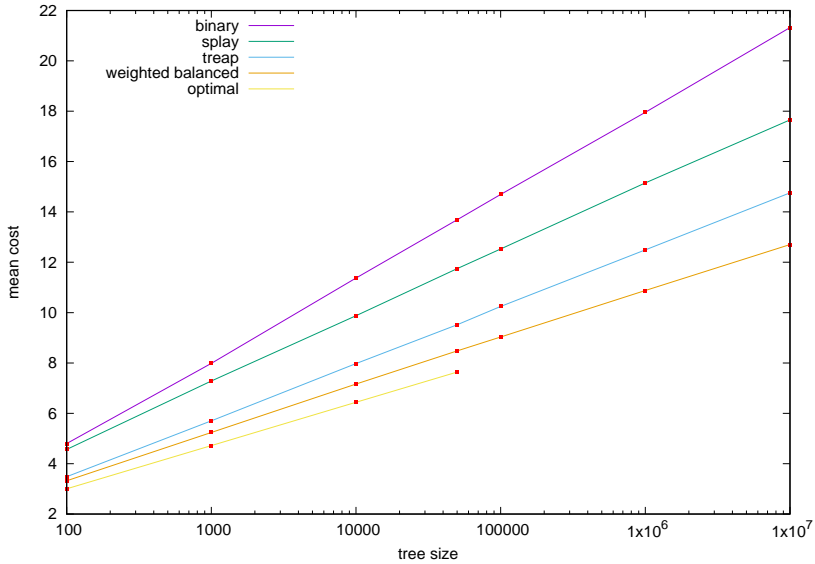}\
\end{center}
\caption{: Baseline performance}
\label{f.base}
\end{figure}

All experiments used 500 samples for each size considered, and the
basic results appear in \fig{base}.  Tree size is on the $x$-axis
using a log scale, and the average number of node expansions during a
single (assumed to be successful) search is on the $y$-axis.

The curves in the legend are ordered from worst to best; simple binary
trees perform the worst (not surprisingly), followed by splay trees,
followed by treaps, and weight-balanced binary trees perform the best.
Expected cost is essentially linear in the log of the size of the
search tree, as one might expect.

\section{Technique}
\label{s.technique}

We now turn to our new idea, which is simply to hill climb using
rotations to improve the expected cost of the tree.

\begin{defn} Let $T$ be a tree.   $T$ will be called {\em weighted\/}
  if there is a function $p:T\rightarrow\R$ that assigns a weight to
  each node in $T$ and for which $\sum_{x\in T} p(x) = 1$.

  If $x\in T$ is not the root of $T$, we will denote the parent of $x$
  by $\pi(x)$.

  We will use $\emptyset$ as a placeholder for a node that is not in
  $T$.  So, for example, we take $\pi(x)$ to be $\emptyset$ if $x$ is
  the root of the tree.

  If $T$ is binary and $x\in T$, we denote the left child of $x$ by
  $l(x)$ and the right child of $x$ by $r(x)$.  Thus $x$ is a fringe
  node if $l(x)=r(x)=\emptyset$.  We denote the sibling of $x$ by
  $\sigma(x)$.

  For a node $x\in T$, we will define the {\em depth\/} of $x$, to be
  denoted by $d(x)$, to be $d(\emptyset)=0$ and $d(x) = 1 + d(\pi(x))$
  otherwise.

  Given a weighted tree $T$, the {\em cost\/} of $T$, to be denoted
  $c(T)$, will be defined to be
  \[c(T)=\sum_{x\in T} d(x)p(x).\]
\end{defn}

One additional property that will be of interest to us is a node that
is either the left child of the left child of its grandparent, or the
right child of the right child of its grandparent.

\begin{defn}  Let $T$ be a binary tree, and $x\in T$ not the root node.
  We will define the {\em like-minded child of $x$}, to be denoted by
  $\lambda(x)$, to be:
  \[\lambda(x) = \cases{l(x), & if $x=l(\pi(x))$; \cr
    r(x) & if $x=r(\pi(x))$.}\]
\end{defn}

Many algorithms manipulate binary trees via what are known as {\em
  rotations}.  As an example, if $T$ is of the form shown in
\fig{left}, where each of $c$, $d$ and $e$ may be the roots of further
subtrees, the result of operating on $T$ with a {\em right rotation\/}
is the new tree shown in \fig{right}.

\begin{figure}[ht]
\begin{center}
  \begin{tikzpicture}[
      every node/.style = {minimum width = 2em, draw, circle},
      level/.style = {sibling distance = 30mm/#1}
    ]
    \node {a}
    child {node {b}
      child {node {c}}
      child {node {d}}
    }
    child {node {e}};
  \end{tikzpicture}
\end{center}
\caption{\ }
\label{f.left}
\end{figure}

\begin{figure}[ht]
\begin{center}
  \begin{tikzpicture}[
      every node/.style = {minimum width = 2em, draw, circle},
      level/.style = {sibling distance = 30mm/#1}
    ]
    \node {b}
    child {node {c}}
    child {node {a}
      child {node {d}}
      child {node {e}}
    };
  \end{tikzpicture}
\end{center}
\caption{\ }
\label{f.right}
\end{figure}

Note that the new tree retains the ordering of the original
(presumably $c<b<d<a<e$), and that $c$, $d$ and $e$ remain apparent
fringe nodes in the new tree so that if they are in fact the roots of
subtrees, those subtrees can remain attached as previously.  The
rotation here is generally referred to as the {\em right rotation
  rooted at $a$,} where $a$ is the root of the subtree being rotated.

Similarly, we will say that a {\em left rotation at $b$\/} produces
\fig{left} from \fig{right}.

\begin{defn} Let $T$ be a tree, and let $x\in T$ be a non-root node.
  By the result of {\em bumping} $x$ in $T$, to be denoted by $T^x$,
  we will mean the result of right rotating $T$ at $\pi(x)$ if $x$ is
  the left child of $\pi(x)$, and the result of left rotating $T$ at
  $\pi(x)$ if $x$ is the right child of $\pi(x)$.  If $r$ is the root
  of $T$, we will take $T^r$ to be $T$ itself.
\end{defn}

Thus \fig{right} shows the result of bumping $b$ in \fig{left}, and
\fig{left} is the result of bumping $a$ in \fig{right}.

Recall that we are assuming that for any particular node $x\in T$, the
weight $p(x)$ is the probability that $x$ is the target of a randomly
selected search query.  The essential point underlying our ideas is
that the impact of bumping $x\in T$ can be computed exactly based
purely on values of a handful of nodes surrounding $x$.

\begin{defn} For a node $x$ in a tree $T$, we will denote by $\overx$
  the subtree of $T$ rooted at $x$.  We will denote by $\overp(x)$ the
  aggregated weight of the points in $\overx$:
  \[\overp(x) = \sum_{y\in\overx} p(y).\]
  We take $\overline{\emptyset} = \emptyset$ (the empty set), and thus
  $\overp(\emptyset) = 0$.
\end{defn}

\begin{prop}  Let $T$ be a weighted binary tree, and $x$ a non-root
  node in $T$.  Then
  \begin{equation}
    C(T) - C(T^x) = p(x) + \overp(\lambda(x)) - p(\pi(x)) - \overp(\sigma(x))
    \label{e.bump}
  \end{equation}
  \label{p.bump}
\end{prop}

\proof This is clear.  Examining Figures \ref{f.left} and
\ref{f.right}, we see that the node $b$ has its depth reduced by 1,
reducing the expected cost of a lookup by $p(b)$; the node $a =
\pi(b)$ has its depth increased by 1.  The entire subtree rooted at
$c=\lambda(b)$ has its depth decreased by 1, reducing the cost by
$\overp(\lambda(b))$ while the subtree rooted at $e=\sigma(b)$ has its
depth increased by 1.  Right bumps are similar, and \eq{bump}
follows.\proofend

\begin{defn} We will refer to the quantity in \eq{bump} as the
  {\em merit\/} of bumping $x$, denoting it by $\mu(x,T)$, or simply
  by $\mu(x)$ if no ambiguity is possible.

  If $T$ is a weighted binary tree, we will say that the {\em result
    of bumping $T$}, to be denoted $\beta(T)$, is:
  \[\beta(T) = \cases{\set T, & if $\mu(x,T)\leq 0$ for all $x\in T$;\cr
    \set{T^x|\mu(x,T)\mbox{is maximal}}, & if $\mu(x,T)>0$ for some $x\in T$.}\]

  For an integer $k\geq 0$, we define $\beta^k(T)$ recursively by
  $\beta^0(T)=\set T$ and
  \[\beta^k(T) = \cup_{t\in \beta^{k-1}(T)} \beta(t)\]
\end{defn}

The notation would be somewhat simpler (but less precise) if we either
assumed there to be a unique $x\in T$ for which $\mu(x)$ was
maximal, or allowed $\beta(T)$ to be the result of bumping $T$ by an
arbitrarily selected $x$ if multiple choices were available.

Since we only bump trees at nodes that have positive merit, we
immediately have:

\begin{lemmx} Let $T$ be a weighted binary tree and $T'$ any element
  of $\beta(T)$.  Then $c(T) \geq c(T')$, with equality if and only if
  $\beta(T) = T$.\proofend
\end{lemmx}

\begin{prop}  Let $T$ be a weighted binary tree.  Then there is some
  finite $k$ such that $\beta^{k+1}(T) = \beta^k(T)$.
\end{prop}

\proof The result will hold for any $k\geq N$, where $N$ is the total
number of binary trees of the given size.  Since each tree has a fixed
associated cost, there can be no sequence of properly increasing costs
(as per the lemma) of size greater than $N$.\proofend

\begin{prop}  Let $T$ be a weighted binary tree of size $n$.  Then
  if $k$ is a positive integer, an element of $\beta^k(T)$ can be
  found in time \[O((n + k)\log(\min(n,k))).\]
  \label{p.cost}
\end{prop}

\proof The result will follow from the following:
\begin{enumerate}
\item It is possible to compute both $\overp(x)$ and $\mu(x)$ for all
  $x\in T$ in time $O(n)$ in an initialization phase, along with 
  \item A list of at least $\min(n,k)$ nodes, sorted by merit, that can be
    initialized in time $O(n\log(\min(n,k)))$,
  \item When a node is bumped, it is possible to update the various
    $\overp(x)$ and $\mu(x)$ in time $O(1)$, and
  \item It is possible to update the merit-sorted list of nodes in
    time $O(\log(\min(n,k)))$.
\end{enumerate}
 The proposition then follows immediately, since the running time is
 $O(n + n\log(\min(n,k)))$ for the initialization and
 $O(1+\log(\min(n,k)))$ for each of $k$ iterations.  The total running
 time is thus as described in the statement of the proposition.

 \begin{enumerate}
 \item $\overp$ can be computed for all of the nodes in the tree in
   time $O(n)$ by simply working back from the fringe.  $\mu$ can then
   be computed, also in time $O(n)$, by virtue of \prp{bump}.
 \item Given merits for all of the nodes, the $k$ highest-merit nodes
   can be found in time $O(n\log k)$.  If $k>n$, it suffices to simply
   sort the merits, so that the time is therefore
   $O(n\log(\min(n,k)))$.
 \item Consider Figures \ref{f.left} and \ref{f.right}.  When a node
   $x$ is bumped, the value of $\overp(x)$ will change only for $x$
   itself, since its parent is now a child, and $x$'s parent $\pi(x)$,
   since both $x$ and one of the subtrees rooted at a child of $x$ are
   no longer descendants of $\pi(x)$.  Computing $\overp(z)$ for each
   of these two nodes takes constant time.

   Similarly, $\mu(z)$ changes only for the nodes labeled $a$ through
   $e$ in the figures.  Again, the update takes constant time.
 \item When the merits are recomputed, inserting any new positive
   values into the list of nodes to bump takes time $O(\log s)$, where
   $s$ is the size of that list, with $s\leq\min(n,k)$ as a result.
   Note that we don't need to search the entire tree for a node to
   ``replace'' the one that was just bumped; if we only plan on
   bumping $k$ nodes in total, we will be interested in one fewer node
   on the next iteration.\proofend
 \end{enumerate}

\section{Experimental results}
\label{s.exp}

It follows from the results of the previous section that we can use our
ideas to reduce the expected cost of searching a binary tree, and that
doing so can be done in a timely fashion.  It is not clear, of course,
whether the methods will be effective in practice.

\begin{figure}
\begin{center}
  \includegraphics[width=\textwidth]{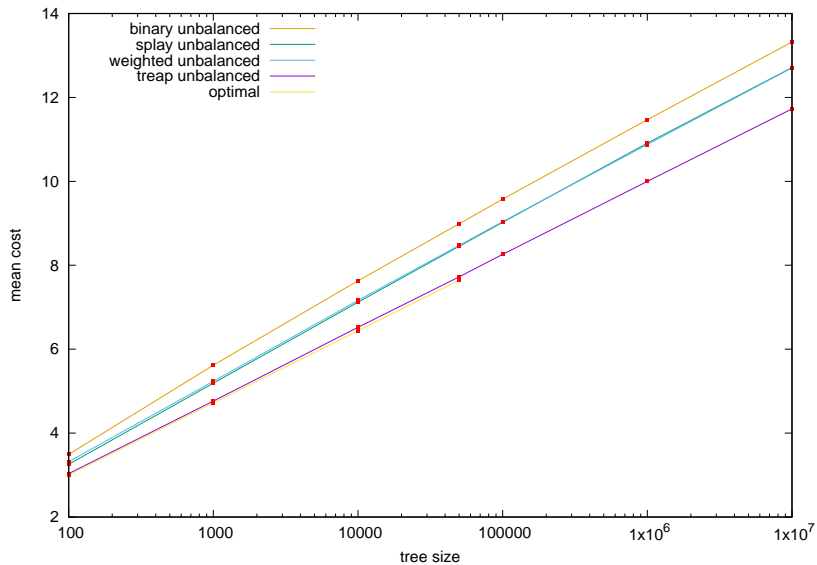}\
\end{center}
\caption{: Algorithm performance after unbalancing the trees}
\label{f.unbalanced}
\end{figure}

For each of the four basic tree constructions described in
\secn{base}, we bumped nodes with the highest merit until quiescence,
so that the tree could not be improved further with our simple method.
The results are shown in \fig{unbalanced}.

Two immediate observations are that the cost remains essentially
linear in the logarithm of the tree size, and that the techniques we
have proposed are by no means a panacea.  They work well for some
trees and clearly less well for others.  Balanced binary trees, for
example, provided the worst performance both before and after the
trees are unbalanced.

Beyond that, however, the results are more interesting.  Unbalanced
splay trees and unbalanced weight-balanced trees perform virtually
identically, even though their performance was starkly different
before unbalancing.  (In actuality, the weight-balanced trees
benefited almost not at all from our ideas; the splay trees improved
substantially.)

\begin{figure}
\begin{center}
  \includegraphics[width=\textwidth]{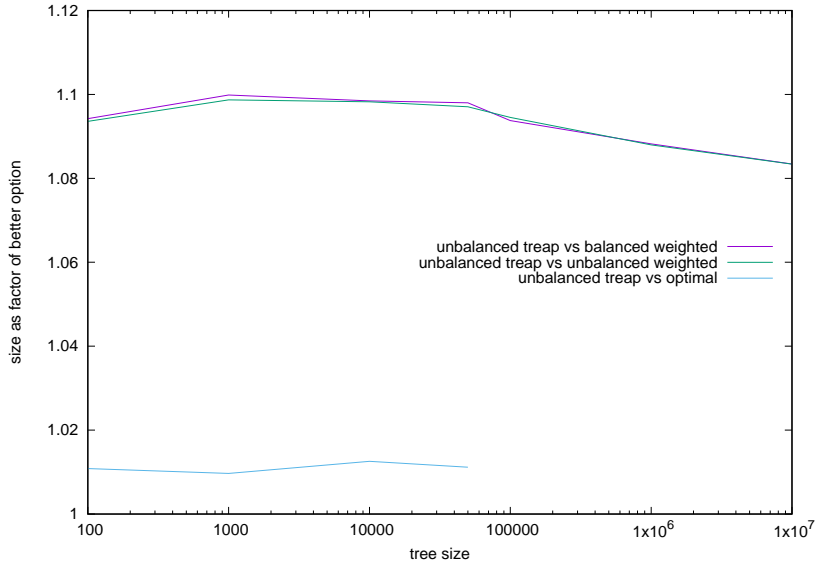}\
\end{center}
\caption{: Comparative performances}
\label{f.factor}
\end{figure}

Perhaps more surprising is that after unbalancing, the treaps have
become easily the most effective method, and their performance is very
nearly optimal for all tree sizes.  As shown in \fig{factor}, the
approximately 9--10\% improvement optimization (relative to either
weight-balanced trees or their unbalanced versions) will provide
computationally meaningful savings in practice; the 1\% difference
between the unbalanced treaps and optimal trees is much less
interesting.

\begin{figure}
\begin{center}
  \includegraphics[width=\textwidth]{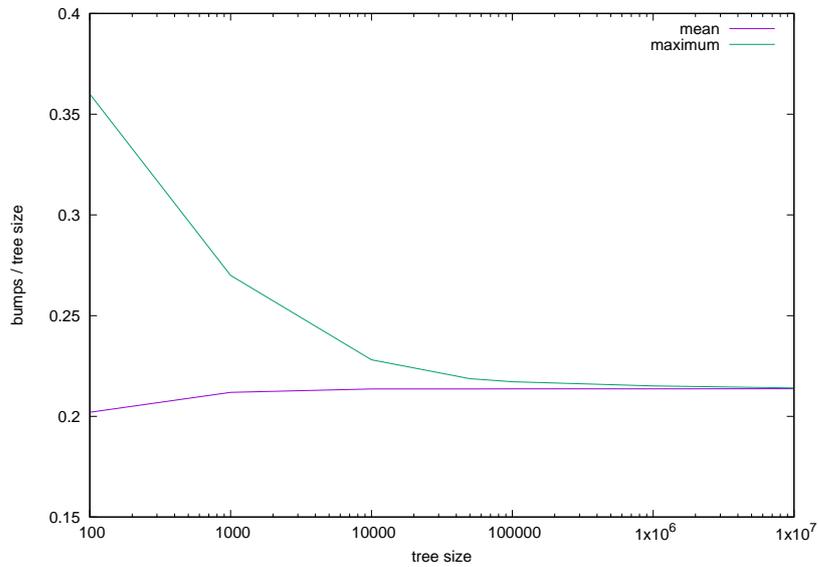}\
\end{center}
\caption{: Mean and maximum number of bumps, as a fraction of tree size}
\label{f.bumps}
\end{figure}

Of course, given the weak bounds on the number of {\em possible\/}
bumps before quiescence, our methods might not be viable in practice
after all.  Data regarding this (for unbalancing treaps, which appears
to be the situation of interest) appears in \fig{bumps}.  For each size
tree, we show both the mean number of bumps improving a treap of that
size, and the maximum number of bumps.  As can be seen, both numbers
stabilize at $b\approx 0.21 n$, where $b$ is the number of bumps.  It
follows from \prp{cost} that the optimized tree can be found in time
$O(n \log(n))$, a complexity identical to the cost of building the
tree in the first place.

\begin{figure}[ht]
\begin{center}
  \begin{tikzpicture}[
      every node/.style = {minimum width = 2em, draw, circle},
      level/.style = {sibling distance = 30mm/#1}
    ]
    \node {0.02}
    child {node {0.49}}
    child {node {0.49}
    };
  \end{tikzpicture}
\end{center}
\caption{: A tree that cannot be locally optimized}
\label{f.bad}
\end{figure}
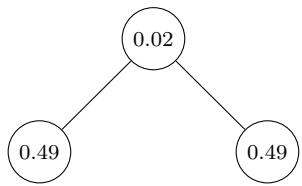

Before concluding, we should expand slightly on our earlier remark
that our ideas are not a panacea; consider the tree in \fig{bad} where
the nodes are labeled with their respective probabilities..

This tree is locally optimal under rotation, since either a left or
right rotation will move one high probability node up and the other
down (for no net benefit), but move the low probability node down for
a net loss.  The optimal tree, which has one high probability node at
the root and the other at depth 1, cannot be reached in a single
rotation from the tree in the figure.

\section{Conclusion and future work}
\label{s.next}

The techniques that we have described appear to lead to clear and
measurable improvements in access times for binary trees, but only
scratch the surface of potential applications.  Some obvious
candidates for future work:

\paragraph{Non-binary trees} There is no reason to restrict our ideas
to binary trees; any data structure where a similar measure of merit
can be computed locally should be amenable to similar treatment.

\paragraph{Unbalancing restrictions} It is possible to apply additional
constraints when selecting a node to bump.  As an example, if we begin
with a binary tree where the maximum node depth is $d$, we could
require that no node can be bumped if it pushes another node below
depth $d$.  This would lead to guaranteed performance improvements
with no impact on worst-case performance.  We could also obviously
limit the maximum depth of a post-bump node in some less restrictive
way.  Since it is possible to compute and maintain $d(\overx)$ for
each node $x$, the impact on computational expense should be
minimal.\footnote{But not zero.  It is possible that a rotation pushes
  a node lower in the search tree, disallowing a future bump that
  would otherwise have been permitted.  This may then require
  extending the heap of best future bumps.}

\begin{figure}
\begin{center}
  \includegraphics[width=\textwidth]{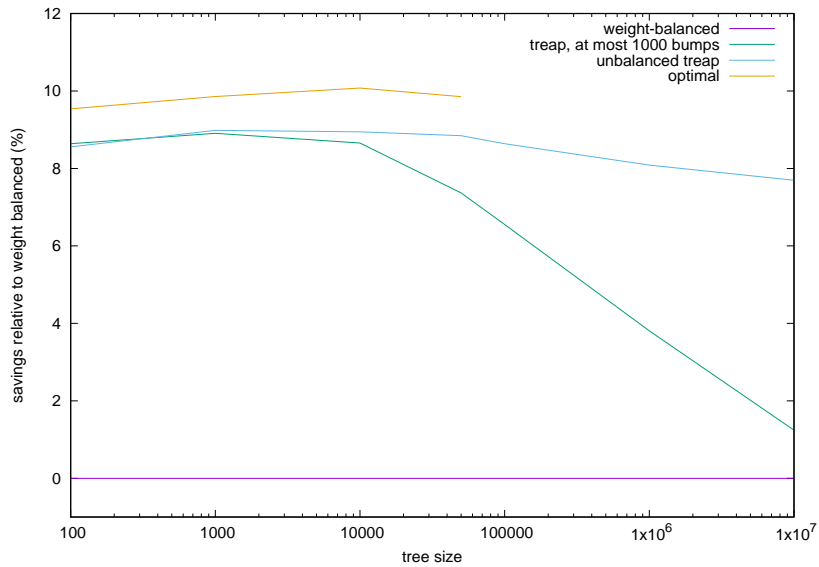}\
\end{center}
\caption{: Impact of limiting the number of bumps}
\label{f.limit}
\end{figure}

\paragraph{Termination before quiescence} Alternatively, one could
limit the number of bumps to be a sublinear function of tree size in
some way, although this may be difficult without significantly
impacting performance.  \fig{limit}, for example, shows performance
both for optimized treaps and for treaps where at most 1000 bumps were
permitted before the optimization was stopped.  As can be seen,
allowing our methods to proceed until quiescence provides significant
improvement without changing the overall complexity of constructing
the tree.  Of course, we have no {\em guarantee\/} that the number of
bumps will be linear in tree size, and we could obviously limit the
number of bumps to (say) 50\% of the number of nodes in the tree.
Given the ``maximum'' curve in \fig{bumps}, doing so would not impact
any of the results we have presented.

\paragraph{Multiple bumps} One can imagine situations where bumping a
node once degrades overall performance, but bumping it twice provides
an improvement.  In general, updating the merits after a single bump
involves recomputing values for 5 nodes; updating the merits after a
bump of $l$ levels will involve recomputing values for $2l+3$ nodes.

It follows that as long as the depth of the tree is bounded at some
fixed multiple of $\log(n)$, we can consider not just single bumps but
bumps of any number of levels and only add a factor of $\log(n)$ to
the worst-case running time.  Of course, given that we are producing
trees with search costs within 1\% of optimal, the additional
complexity of such considerations may not be warranted.

Testing all of these ideas should be straightforward.

\section*{Acknowledgment}

I am grateful to Don Knuth, Cecilia Aragon, and an anonymous reviewer
for a variety of extremely insightful comments and suggestions.

\section*{Conflict of interest}

\paragraph{Funding:} Not applicable.

\paragraph{Conflicts or competing interest:} None.

\paragraph{Availability of data, material and code:} Code and other
materials will be provided upon request.  The author reserves the
right to patent the techniques described here.  In the event that such
patents are filed and granted, publication of this paper carries with
it a permanent, royalty-free license for noncommercial use of the
technology herein described.

\bibliography{ubt}

\end{document}